    \documentstyle[epsf]{article} \catcode`\@=11
    \def\section{\@startsection{section}{1}{\z@}%
    {-3.5ex plus -1ex minus -.5ex}{1.5ex plus.3ex}{\bf }}
    \def\subsection{\@startsection{subsection}{1}{\z@}%
    {-3.5ex plus-1ex minus-.5ex}{1.5ex plus.3ex}{\bf }} 
\def\BRA{\left\langle}
\def\KET{\right\rangle}

\def\GP{G^{+}}
    
\begin{document}
{\Large\bf  S-matrix network models \\ for coherent waves in random media:
construction and renormalization
    }\vspace{.4cm}\newline{\bf   
Martin Janssen, Rainer Merkt and Andreas Weymer
    }\vspace{.4cm}\newline\small
Institut f\"ur Theoretische Physik, Universit\"at zu
K\"oln, Z\"ulpicher Strasse 77, 50937 K\"oln, Germany
    \vspace{.2cm}\newline 
  6 October 1998 
    \vspace{.4cm}\newline\begin{minipage}[h]{\textwidth}\baselineskip=10pt
    {\bf  Abstract.}
Networks of random quantum scatterers (S-matrices) form
paradigmatic models for the propagation of coherent waves in random
media. S-matrix network models cover universal localization-delocalization
properties and have some advantages over more traditional Hamiltonian
models. In particular, a straightforward implementation of real space
renormalization techniques is possible.  Starting from a finite
elementary cell of the S-matrix network, hierarchical network models can be
constructed by recursion. The localization-delocalization properties
are contained in the flow of the forward scattering strength
('conductance') under increasing system size. With the aid of 'small
scale' numerics qualitative aspects of the localization-delocalization
properties of S-matrix network models can be worked out. 
    \end{minipage}\vspace{.4cm} \newline {\bf  Keywords:}
 S-matrix network models, Localization-delocalization transition,
renormalization group, multifractality
    \newline\vspace{.2cm} \normalsize
\section{Introduction}\label{sec1}
Localization and fluctuation phenomena are fascinating universal
features in any disordered coherent wave mechanical system.  In 1982
Shapiro \cite{Sha82} pointed out that a convenient modeling of
disordered coherent wave mechanical systems can be given by networks
of unitary random scattering matrices. This modeling covers essential
symmetries and characteristic length scales, but does not rely on
particular dispersion relations and specific details. Therefore, it is
of relevance in various fields of theoretical physics, e.g.  in
optics, mesoscopic electronics, and quantum chaos.  S-matrix network
models (NWMs) have a number of advantages over more traditional
Hamiltonian models. NWMs yield directly transport quantities
\cite{Met98}, wave packet dynamics \cite{Huc98}, and quasi-energy
eigenvalues and eigenstates \cite{Kle97}.  A well known NWM introduced
by Chalker and Coddington (CC-model) \cite{Cha88} (see also
\cite{Fer88}) describes the situation of disordered two-dimensional
(2D) electrons undergoing the quantum Hall localization-delocalization (LD)
 transition in a strong
perpendicular magnetic field. A major advantage of S-matrix NWMs
is a straight forward implementation of  real space renormalization
group (RG) techniques \cite{Gal97Aro97,Wey98} which open new
perspectives for investigating LD
transitions.

Here we address the construction of S-matrix NWMs starting from
symmetry considerations and/or from an Anderson tight-binding
Hamiltonian.  We discuss alternative implementations for a RG action.
To reach a qualitative overview of phases in the phase space of
scattering strengths one can construct hierarchical network models for
which a RG algorithm can be formulated that is designed for 'small
scale' numerics. The construction of S-matrix NWMs and the RG methods
are demonstrated on particular two-dimensional models.
\section{Construction of S-matrix networks}\label{sec2}
Quite generally, a NWM can be constructed as follows. Take a regular
network of ${\cal N}$ sites and $N$ bonds.  Each bond $\alpha$ carries
propagating wave modes ($n^{+}_\alpha$ incoming modes and
$n^{-}_\alpha$ outgoing modes) represented by complex amplitudes,
$\psi_{n^{\pm}_\alpha}$.  On the sites unitary S-matrices map incoming
to outgoing amplitudes.  The elements of each S-matrix are (in
general) random quantities respecting the symmetries of S and are
characterized by typical scattering strengths.  Random phases are
attached to the amplitudes on the bonds. They simulate the random
distances between scatterers in realistic systems.
\subsection{Topology and  site S-matrix}\label{sec2.1}
The construction of a NWM is fixed by the choice of a certain type of
random S-matrix 
\begin{figure}[b] 
\epsfysize=3.5cm
\centerline{\epsfbox{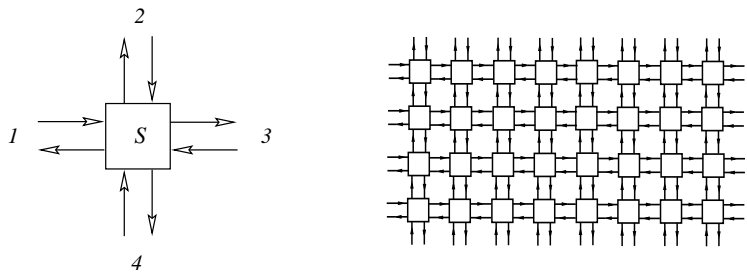}}
\caption{\protect\small U2 network model with unitary
S-matrices. On the left the elementary S-matrix and on the
right a section of the whole network are graphically represented.}
\label{JMW98FIG1}
\end{figure}
and a ${\cal N}\times {\cal N}$ connectivity matrix $C$ which has
elements $C_{ij}=1$ if a wave mode can propagate from site $i$ to site
$j$ and $C_{ij}=0$ otherwise.  The NWM defines a dynamical system
by $S$ and $C$. Together they lead  to a unitary matrix $U$ \cite{Fer88,
Kle95,Edr88} called 'network operator'. It maps all incoming to
outgoing wave modes. We choose a convenient time unit and denote the
vector of the $B\propto N$ bond amplitudes as $\psi$ such that
$\psi(t+1)=U \psi(t)$.  The eigenphases $\phi_n$ ($n=1\ldots B$) and
corresponding eigenvectors $\psi_n$ of $U$, $ U\psi_n = \exp(i\phi_n)
\psi_n \, , $ can then be interpreted as quasi-energies and
corresponding eigenstates.
\subsection{Two-dimensional networks}\label{sec2.2}
Let us now report on two-dimensional NWMs with the topology of a
square lattice and a connectivity matrix that connects each site with
its four nearest neighbors. The corresponding S-matrix is
graphically represented in Fig.~1 and reads 
\begin{equation}
	S=\left( \begin{array}{cccc} r^1 & d_{R}^{21} & t^{31} & d_{L}^{41} \\
	d_{L}^{12} & r^2 & d_{R}^{32} & t^{42}\\
	t^{13} & d_{L}^{23} & r^3 & d_{R}^{43}\\
	d_{R}^{14} & t^{24} & d_{L}^{34} & r^4 
	\end{array} \right)\, . \label{SM}
\end{equation}
The parameters of each S-matrix are the amplitudes of transmission
$t$, reflection $r$, and deflection to the left $d_L$ and right $d_R$
which contain random phases compatible with symmetry requirements.
Scattering strengths of a site S-matrix allow to define transport mean
free paths $l_e$ (measured in
units of the lattice constant) for a prescribed forward direction  
by $ 1/l_e = (1-T)/T \, .  $ Here $T$
is the probability to be transmitted in the prescribed forward
direction.  Such definition is known from the scattering theoretical
approach to quasi-1D systems (e.g.  \cite{Mac92}).

The unitarity of the S-matrix requires $|t|^2+|r|^2+
|d_L|^2+|d_R|^2=1$.  The corresponding parameter space of scattering
probabilities, a quarter of a pyramid, is shown in Fig.~2 and forms
the phase space of the U2-model ('U' stands for unitary, '2' for the
two-dimensional lattice structure).
\begin{figure}[h] 
\epsfysize=5cm
\centerline{\epsfbox{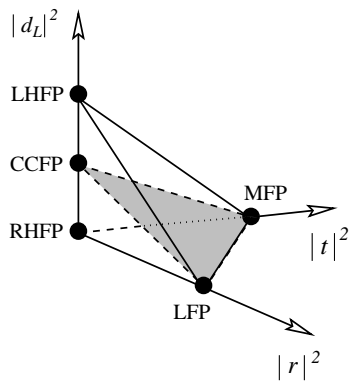}}
\caption{\protect\small The three-dimensional 
phase space of the U2-model is spanned by
scattering strengths $0< |t|^2,|r|^2,|d_L|^2 < 1$ with
$|t|^2+|r|^2+|d_L|^2= 1-|d_R|^2$ and $|d_R|^2<1$. The shaded area
corresponds to the U2NC-model for which $|d_L|=|d_R|$. The points
LHFP, RHFP, LFP and MFP are characterized by $|d_L|=1$, $|d_R|=1$,
$|r|=1$ and $|t|=1$, respectively. They form the corners of the U2-phase
space. The critical fixed point of the Chalker-Coddington model
(CCFP) corresponds to $|d_L|^2=|d_R|^2=1/2$.}
\label{JMW98FIG2}
\end{figure}
  The points in the corners of the phase space are the metallic fixed point
(MFP) ($|t|=1$), the localization fixed point (LFP) ($|r|=1$), and the
left (right) handed fixed point LHFP (RHFP), $|d_L|^2=1$ ($|d_R|^2=1$).
The term 'fixed point' refers to the renormalization group action to
be discussed below. 
\subsection{Symmetries and phase space}\label{sec2.3}
In the absence of handedness of the scattering process,
i.e. $|d_L|=|d_R|$, the model is denoted as non-chiral (U2NC-model).
The phase space of the U2NC-model corresponds to the shaded plane in
Fig.~2. The point of $|d_L|^2=|d_R|^2=1/2$ is denoted as
Chalker-Coddington fixed point (CCFP) since the U2-model reduces to
two uncoupled CC-models for vanishing $t$ and $r$ and the phase space
of the CC-model is one-dimensional (1D) and parameterized by the values
of $T:=|d_L|^2$ (see Fig.~3a).
 
Further symmetries of the S-matrix lead to a reduction of the number
of independent parameters. This typically eliminates random phases,
but can also lead to restrictions in the phase space of scattering
strengths.  For time reversal symmetric scattering in the U2NC-model
the S-matrix is symmetric and the phase space is reduced by
$|r|+|t|\geq 1$. The model is then denoted as O2NC-model where 'O'
stands for orthogonal (the network operator can be diagonalized by
orthogonal matrices).
\begin{figure}[h] 
\epsfysize=5cm
\centerline{\epsfbox{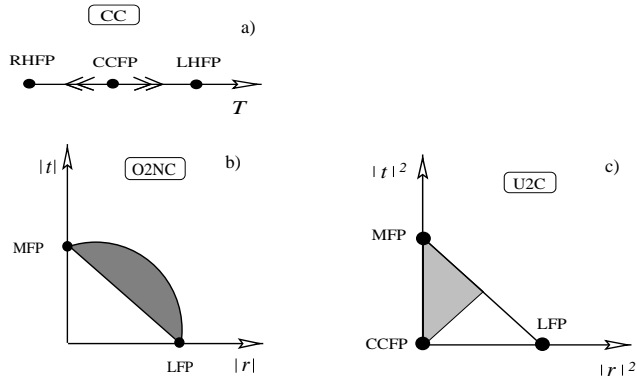}}
\caption{\protect\small a) The phase space of the Chalker-Coddington
model. Two of the fixed points correspond to quantum Hall
plateaus (LHFP and RHFP),  the critical point CCFP corresponds to
the plateau to plateau transition. b) The phase space of the
O2NC-model in $|r|,|t|$ parameter space. The time reversal symmetry
leads to the restriction $|t|+|r|\geq 1$ (grey area). c) The phase
space of a chiral model. The chiral symmetry (see text) 
leads to the restriction
$|t|^2\geq|r|^2$  (grey area).}
\label{JMW98FIG3}
\end{figure}
Its phase space is shown in Fig.~3b. Imposing symmetry constraints
reflecting a sublattice structure, e.g.
$\Sigma_3S\Sigma_3=S^\dagger$, where $\Sigma_3:={\rm diag}(1,1,-1,-1)$,
the U2NC-model becomes '$\Sigma_3$-chiral'.  Although $|d_L|=|d_R|$ is
still valid, the chiral symmetry leads to further phase space
constraints, e.g. to $|r|^2\leq |t|^2$ (for $d\not=0$) for
$\Sigma_3$-chirality (Fig.~3c).
\subsection{From tight binding models to S-matrix networks}\label{sec2.4}
So far S-matrix network models appeared as models on their own, dictated by
topology and symmetry. However they can also appear as the result of a
mapping from a microscopic Hamiltonian, as is the case for the
CC-model \cite{Fer88}. In this subsection we show how tight
binding models can also lead to S-matrix network models. This opens a
new flexibility in the methods to study these models.

The general idea is to take a typical microscopic cell of a given tight
binding model 
and attach 1D semi-infinite leads to it, as many as is
appropriate for topology and  symmetry
aspects. Then calculate the S-matrix of this device. It 
may then serve to define a NWM.

We proceed as follows. The Hamiltonian $H=H_0+H_1$ is decomposed into
a part that describes only the leads, $H_0$, and the part that
describes the microscopic cell together with its coupling to the
leads, $H_1$.
The scattering states with
continuous spectrum are chosen to obey energy normalization $\BRA
\alpha(E)|\alpha(E')\KET=\delta (E-E')$. Then the S-matrix is given as
$
	S_{\alpha \beta}=\delta_{\alpha \beta} -4\pi i T_{\alpha \beta}
$
where the T-matrix (operator) is related to the resolvent (Green's
function)
 $\GP=(E+i0^{+}-H)^{-1}$ by
$
	T=H_1 + H_1\GP H_1 \, .
$
Note that  $H_0$  describes 'isolated' terminals, i.e. 
(semi-infinite leads) and  the modes $\left.\mid \alpha \KET$ contain
both `in`- and `out`-states. Introducing the
projector $Q=1-P$  onto the
Hilbert-space of the terminal states the T-matrix elements read
\begin{equation}
	T_{\alpha\beta} =\BRA \beta\mid H_{QP} \GP_{PP}H_{PQ}\mid
	\alpha\KET\, .\label{2.1}
\end{equation}
The projected Green's function $\GP_{PP}$ can be determined by a non-hermitian
effective Hamiltonian defined entirely on the $P$-space, i.e. we have
to invert a $V$ dimensional matrix when $V$ is the number of sites in
the microscopic cell.

It is convenient to treat the terminals as  semi-infinite 1D
tight-binding models with only kinetic energy and unit hopping
strength.  The terminal Green's function as well as the amplitudes of
the terminal states can be
calculated analytically.  Together with the solution of the $V\times
V$ matrix problem the S-matrix can be obtained from Eq.~(\ref{2.1}).

To make the discussion more explicit we  present the
 results for two models \cite{Jan-RF}. First consider the
microscopic cell in Fig.~4a consisting of a  on-site random energy
$\varepsilon$
attached to four terminals. For energies
in the band center of the 1D terminals  the scattering matrix elements
are \cite{NOTE1}
$$
	t=d_L=d_R=\frac{8-2i\varepsilon}{16 +\varepsilon^2}
$$  and $r=1-t$.
The model is a member of the O2NC-model since $S$ is symmetric. To get
scattering strengths characteristic for an ensemble of disorder
realizations  one should   average them  with respect to an appropriate
distribution of $\varepsilon$. 
\begin{figure}[h] 
\epsfysize=3.5cm
\centerline{\epsfbox{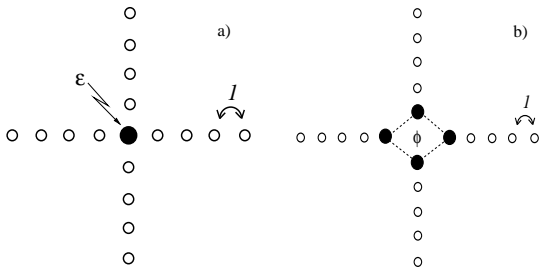}}
\caption{\protect\small Elementary S-matrix for tight binding
Hamiltonian models: a) A central site with random on site energy
$\varepsilon$ is coupled to four semi-infinite leads (terminals)
characterized by unit hopping amplitudes. b) Four central sites
enclosing a random flux $\phi$  are coupled to semi-infinite
leads.}
\label{JMW98FIG4}
\end{figure}

Second we address a microscopic cell for the random flux problem as
shown in Fig.~4b. No on-site disorder is introduced, but a random flux
phase $\phi$ (the flux quantum is set to $2\pi$) is picked up when
moving around the four inner sites (anti-clockwise). At the band
center the resulting scattering amplitudes (for an incoming wave from
the left in Fig.~4b) are found to be
\begin{eqnarray}
	 r&=&(4\sin^2(\phi/2)-1)/D \, ,\;  t
	= -4\cos(\phi/2)/D \, ,\nonumber\\
 	d_L^{12}(\phi)&=& e^{-i\phi/4}(2i e^{i\phi}
	-4i)/D = - (d_R^{21})^\ast \nonumber
\end{eqnarray}
 with $D=4\sin^2(\phi/2) +5$.  A close
inspection shows that the model is '$\sigma_3$-chiral', i.e.
 $\sigma_3S\sigma_3=S^\dagger$ with
 $\sigma_3:={\rm diag}(1,-1,1,-1)$. This property is
lost for  energies  not exactly at the band center.
\section{Implementation of a renormalization group}\label{sec3}
To detect  localized, critical and delocalized phases of
an infinite NWM one likes to implement a convenient renormalization
group (RG). Following the scaling theory of Abrahams et
al. \cite{Abr79} one could think of computing the two-probe
conductance $g(L)$ for a network of size $L^2$
(Fig.~5a).  
\begin{figure}[h] 
\epsfysize=5.cm
\centerline{\epsfbox{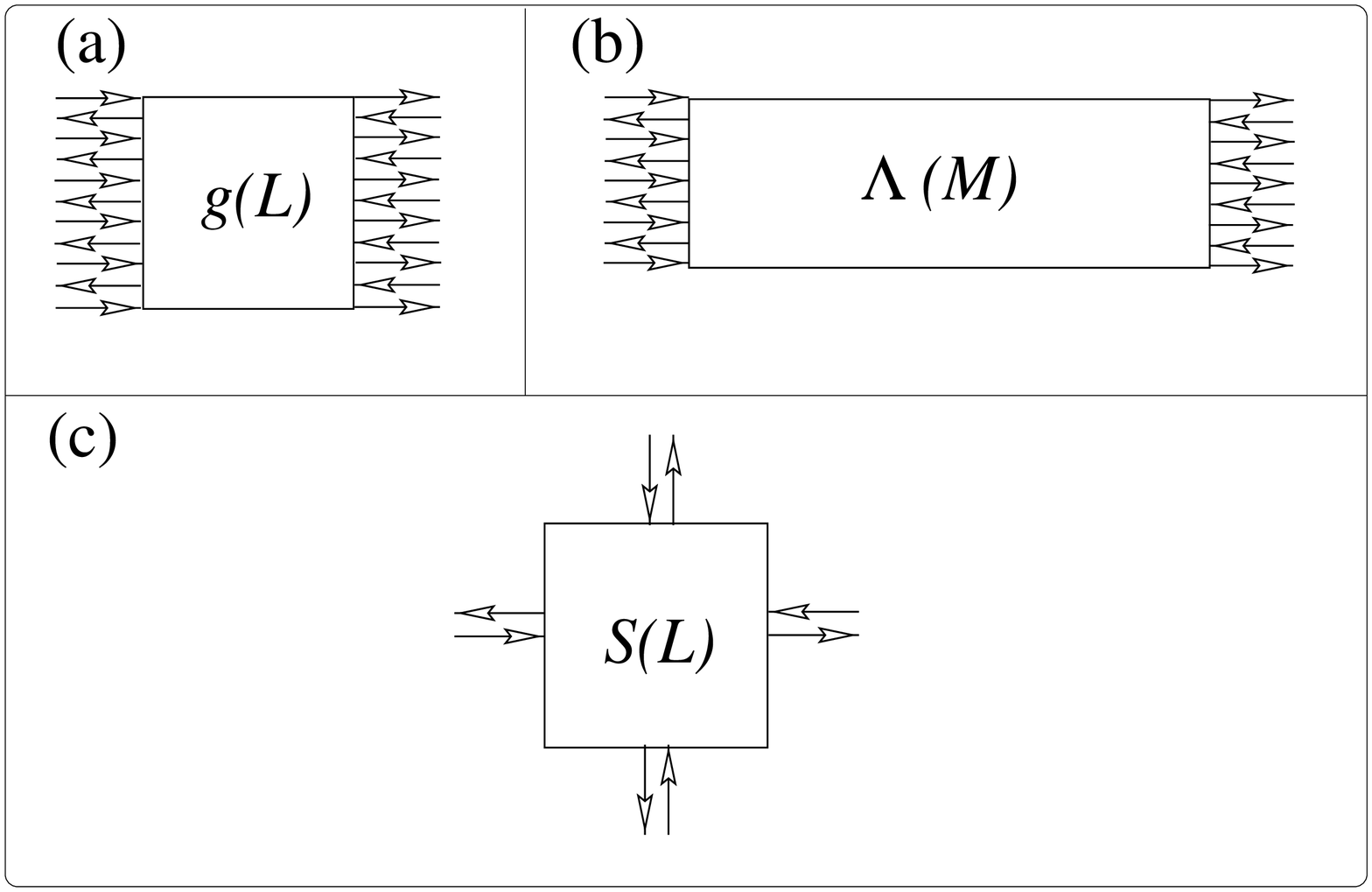}}
\caption{\protect\small Different ways of implementing a
renormalization group: a) The two probe conductance
 $g(L)$ is studied for increasing system size $L$. b) The finite size 
scaling variable
$\Lambda(M)$  is studied as a function of width $M$ in  quasi-1D
geometries. c) The elementary S-matrix can be generalized to an
 S-matrix $S(L)$ for arbitrary system size $L$. }
\label{JMW98FIG5}
\end{figure}
This involves strong mesoscopic fluctuations in $g(L)$. For that
reason the self-averaging object $\Lambda(M)=\xi(M)/M$ has been
introduced \cite{MacK81Pich81} to study the scaling behavior. Here
$\xi(M)$ is the localization length corresponding to quasi-1D
geometries (Fig.~5b) of length $L\to \infty$ and finite width
$M$. This so called 'finite size scaling' method is the 'first choice'
method for numerical investigations of scaling properties
\cite{Kra93}. It has been successfully applied to the CC-model
\cite{Cha88,Kiv93} and to the U2NC-model (and variants of it)
\cite{Fre98Mer98}.  Still, it involves 'large scale' numerics,
i.e. powerful computation units and/or large storage capacities. One
reason for this is that the finite size scaling is an 'exact'
numerical method -- no approximations in the modeling are involved.

Alternatively, one can treat the whole system of linear size $L^2$ as
a big scattering unit (see Fig.~5c) giving rise to a RG flow of the
S-matrix: $S_{\alpha \beta}(1) \longrightarrow
S_{\alpha \beta}(L)$. Consequently, this RG procedure allows to study
a RG flow in the phase space of scattering strengths. In particular
the fixed points of this RG flow are of interest. The repulsive or
attractive behavior of the RG flow close to the fixed points gives
valuable information about phase transitions. For example, the CCFP is
repulsive within the phase space of the CC-model while LHFP and RHFP
are attractive fixed points (see Fig~3a). These simple facts 
contain essentials of the quantum Hall effect.
\section{Hierarchical S-matrix networks}\label{sec4}
The numerical effort in computing a single S-matrix element of $S(L)$
is of the same order as in computing a single eigenstate of the
network operator $U$. Furthermore, one needs computations for
different disorder realizations to take large fluctuations into
account. Thus, a further idea is needed to make advantage of the RG
flow in phase space.  We give up the 'exactness' of the RG method and
consider simplified NWMs by recursive constructions of 'hierarchical'
NWMs as shown in Fig.~6 for the U2-model.
\begin{figure}[t] 
\epsfxsize=4.cm
\centerline{\epsfbox{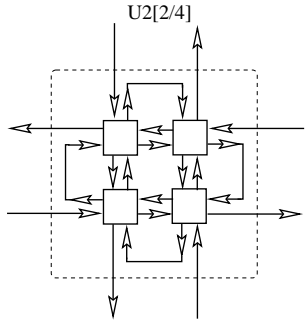}}
\caption{\protect\small An example (denoted as U2[2/4]) for a
 recursive construction of a hierarchical
network model. Four elementary S-matrices are connected such that the
resulting $4\times 4$ network has the same terminal structure as each
of the elementary S-matrices.
}
\label{JMW98FIG6}
\end{figure}
A section of a given NWM  containing $b^2$ sites and $V$ 'active'
elementary S-matrices (S-matrices with only trivial elements $0,1$ are
not counted as active) is taken. The scatterers are connected such that
the resulting network has the same terminal structure as each of the
elementary S-matrices.  Accordingly, the hierarchical models are
denoted as [b/V]-models. The hierarchical models have fractal
dimension $D^{[b/V]}=\ln V/\ln b$.  In the $V \to \infty$ limit the
original NWM may be recovered exactly (depending on the construction).
Obviously, the RG flow works by recursion and the computational effort
simplifies drastically. The
hierarchical NWMs approximate the original ones in an uncontrolled
manner. Better approximations can only be gained by increasing the
size of the elementary cell and by studying the convergence of
quantitative results, e.g.  critical exponents. However, qualitatively
correct results for the existence or non-existence of phase
transitions can be obtained already for moderate sizes of the
elementary cells. The RG method for hierarchical
NWMs yields qualitative  aspects of the RG flow in
phase space of the original NWM relying on  small scale numerics.
\section{Renormalization algorithm}\label{sec5}
In the following we outline the six-step algorithm of the RG in
hierarchical network models.

(i) In the first step $S_{\alpha \beta} (L=b)$ is determined as a
 function of the scattering matrix elements of all active scatterers
 for a given elementary cell (respecting unitarity constraints).  For
 small numbers $V$ this can be done by hand or computer algebra.

(ii) The set of scattering strengths ${\bf T}$ is  drawn from an initial
distribution,$P_0({\bf T})$. 
Phases  are taken uncorrelated
and uniformly distributed in the interval $[0,2\pi]$.  For each
realization ${\bf T}(L=b)$ is calculated with the help of the
algebraic expression obtained in the first step.  

(iii) After collecting a large number ($\sim 10^3$) of scattering
strengths an approximate distribution function $P_{L=b}({\bf T})$ is
determined which depends on the initial distribution $P_0({\bf
T})$. In particular, it yields the average value $\BRA {\bf
T}\KET(L=b)$ and other characteristic parameters of the distribution.

(iv) Steps (ii)  and (iii) are
repeated, but now ${\bf T}$ is drawn from the distribution
$P_{L=b}({\bf T})$ obtained before.  This leads to a reasonable
approximation for $P_{L=b^2}({\bf T})$ and for $\BRA {\bf T}\KET
(L=b^2)$.  

(v)  Step (iv) is repeated $N$ times ($N\sim 10^1$) which
yields, for each initial distribution $P_0({\bf T})$, the flow of the
distribution function $P_{L}({\bf T})$ and of the mean $\BRA {\bf
T}\KET (L)$ as functions of $\ln L=(N+2)\ln b$.  

(vi) Finally, the RG
process ((i) -- (v))  will be repeated for a number of different initial
distributions $P_0({\bf T})$ to study the RG flow of typical values
close to the fixed points.
\section{Small scale numerics}\label{sec6}
RG calculations following this algorithm
 have been performed for
the CC-model \cite{Wey98} (see also \cite{Gal97Aro97}) and the
Manhattan model (MM) considered in  \cite{Zirn97}. The MM 
is defined by $r=0$ and an alternating
structure of $d_L=0$ or $d_R=0$ (see Fig.~8) leaving the forward transmission
$T:=|t|^2$ as the single parameter for its phase space.

The results for a number of CC[b/V]-models \cite{Wey98} can be
summarized as follows: A LD transition occurs, characterized by a
repulsive fixed point distribution $P^{\ast}(T)$ (here $T:=|d_L|^2$)
which is broad over the interval $[0,1]$. After a few iteration steps
the flow becomes independent of the initial distribution and can be
characterized by the average value $\BRA T\KET$. The critical
distributions were found to be symmetric around the critical mean
value $T^\ast=0.5$ and developed a shallow minimum there.  The
critical distribution for the CC[3/5]-model and the scaling of its
mean value is shown in Fig.~7.
\begin{figure}[b] 
\epsfysize=4.5cm
\centerline{\epsfbox{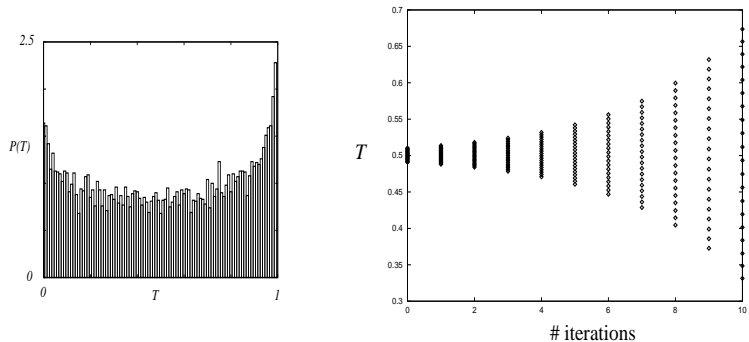}}
\caption{\protect\small  The critical distribution of scattering
strengths (left) and the corresponding flow of average
transmission strengths $T$ under increasing number  of 
renormalization steps (right) for the CC[3/5]-model.}
\label{JMW98FIG7}
\end{figure}
Furthermore, one observes that with increasing elementary cells the
critical exponent  of the localization length
approaches the value of the original CC-network  in a monotonic
way from above. 

In addition to the localization length exponent a second critical
exponent is required to characterize the LD transition, the critical
exponent, $\alpha_0-D$, of the order parameter (the typical density of
states) \cite{JanR98}.  This quantity scales to zero at the critical
point,
$$
	\rho_{\rm t} (L) \propto L^{-(\alpha_0-D)} \; , \; \, \alpha_0>D\, ,
$$
since critical states are not totally space-filling, but multifractal.
Here $D$ is the dimension of the system which may be fractal in
hierarchical NWMs.  Adopting the method of Klesse and Metzler
\cite{Kle95} to calculate wave functions as eigenstates of the network
operator $U$ the multifractal exponent $\alpha_0$ can be computed for
a given network model. To reduce the computational effort, all
scattering strengths can be fixed to the critical value $T^\ast$.
Moderate sizes ($L \approx 50$) are used and several realizations to
obtain $\alpha_0$.  It turns out that $\alpha_0-D^{[b/V]}$
monotonically approaches the value of the CC-model \cite{Kle95} from
above as $D^{[b/V]}$ approaches $d=2$ from below.

It is believed that a crucial ingredient of the CC-model is its
handedness
reminiscent of the strong magnetic field in quantum Hall systems. To
check whether the handedness is essential for the occurrence of the LD
transition we performed a RG process for the Manhattan model which has
a one-dimensional phase space like the CC-model, but no handedness
(even for strong deflection there are as many right circulating loops
as left circulating loops in a large network). Indeed, we find no
indication for a LD transition in the MM. Instead,  any initial
distribution of the forward scattering strength $T$ flows to the strong
localization fixed point LFP (see Fig.~8). 
\begin{figure}[t] 
\epsfysize=9cm
\centerline{\epsfbox{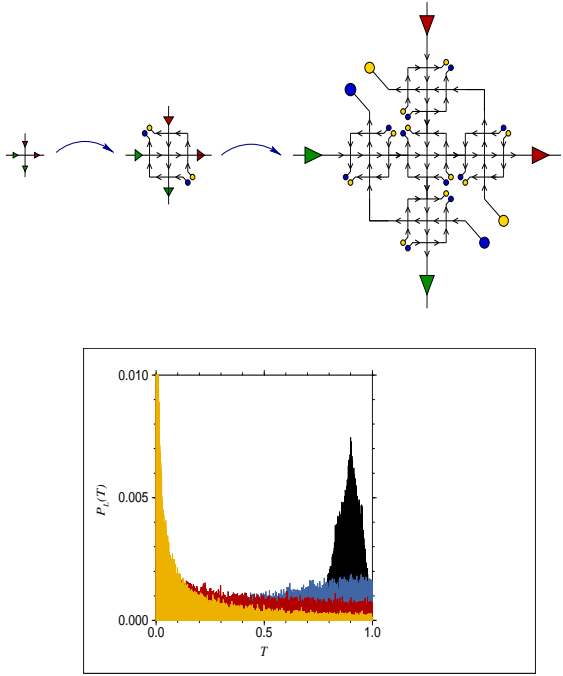}}
\caption{\protect\small The [3/5]- model corresponding to the
Manhattan model (top) and the flow of the distribution of the forward
scattering strength (bottom) under renormalization ($L=3,9,27,81$) to
strong localization. Amplitudes on open ends with the same symbol
(bullet or shaded circle) are identified within one RG step.}
\label{JMW98FIG8}
\end{figure}
\section{Conclusions}\label{sec8}
S-matrix network models provide a flexible tool to study disordered
coherent wave mechanical systems. The construction can be based on
topology and symmetry requirements, but also on microscopic models
such as tight binding Hamiltonians.  For a given network model a
recursive construction of associated hierarchical network models can
be performed.  Iterating this construction yields an approximation of
the original network model which allows to study the flow of
scattering strengths in the phase space of the network model by a six step
renormalization group algorithm.  The precision of critical parameters
obtained by this procedure can be improved by increasing the number of
realizations in each iteration step.  The critical parameters are then
more precise for the particular hierarchical model, however have
uncontrolled deviations from that of the original model. Better
approximations can only be gained by increasing the size of the
elementary cell and studying the convergence of critical exponents.
The computational effort then drastically increases and becomes
comparable to that of the well known finite size scaling method based
on the transfer matrix technique.
The advantage of the  RG method for hierarchical
network models is that it  yields qualitative  aspects of the RG flow in
phase space of the original model using only  small scale numerics.
%

\vspace{0.6cm}
We thank A. Altland, P. Jacquod for helpful discussions and
B. Shapiro for initiating our interest in the RG approach to NWMs.
This work was supported by the SFB 341 of the Deutsche
Forschungsgemeinschaft.  


\begin{thebibliography}{99}
\bibitem{NOTE1} For a two lead configuration the procedure defines a
mean free path which is  consistent with the one resulting from
independent calculations of 1D localization lengths $\xi=2 l_e$
carried out for weak disorder (cf.\cite{Kra93}).
\end{thebibliography}
    \end{document}